\documentclass{article}
\usepackage{epsfig}
\usepackage{color}
\usepackage{amsmath}
\usepackage{amssymb}
\usepackage{lscape}
\usepackage{cite}

\makeatletter
\@addtoreset{equation}{section}

\begin{document}

\begin{flushright}
May 2016

OCU-PHYS 449
\end{flushright}

\begin{center}

\vspace{3cm}

{\LARGE 
\begin{center}
Notes on Planar Resolvents 

of 

Chern-Simons-matter Matrix Models
\end{center}
}

\vspace{2cm}

Takao Suyama \footnote{e-mail address : suyama@sci.osaka-cu.ac.jp}

\vspace{1cm}

{\it 
Osaka City University
Advanced Mathematical Institute (OCAMI)}\\
{\it 3-3-138, Sugimoto, Sumiyoshi, Osaka 558-8585, Japan
}

\vspace{2cm}

{\bf Abstract} 

\end{center}

We revisit planar resolvents of matrix models corresponding to ${\cal N}\ge3$ Chern-Simons-matter theories with the gauge groups of the form ${\rm U}(N_1)\times{\rm U}(N_2)$ coupled to any number of bi-fundamental hypermultiplets. 
We find that the derivative of a suitably defined planar resolvent can be written explicitly. 
From this resolvent, we derive the explicit formula for (a linear combination of) the vevs of BPS Wilson loops. 
As a non-trivial check, we show that the formula reproduces the perturbative expansion of  the vevs of the BPS Wilson loops. 

\newpage

\vspace{1cm}

\section{Introduction}

\vspace{5mm}

AdS/CFT correspondence \cite{Maldacena:1997re} provides us with the possibility to investigate a quantum theory of gravity in terms of an ordinary quantum field theory. 
In order to study some geometrical properties of the dual gravity theory, one is typically required to perform a quantitative analysis on strong coupling behaviors of the corresponding quantum field theory. 
Recently, such an analysis has become manageable, at least for a set of physical quantities, thanks to the developments of the supersymmetric localization. 
See \cite{Pestun:2014mja}\cite{Hosomichi:2014hja} for recent reviews on this topic. 

ABJM theory \cite{Aharony:2008ug} provides a prototypical example of AdS$_4$/CFT$_3$ correspondence. 
This is a Chern-Simons theory coupled to matters with ${\cal N}=6$ superconformal symmetry. 
The analysis of this theory has been done intensively in the context of AdS/CFT correspondence as well as in relation to M2-branes. 
The researches discussing the strong coupling behaviors of ABJM theory include the ones using the planar limit \cite{Suyama:2009pd,Marino:2009jd,Drukker:2010nc,Drukker:2011zy}, the M-theory limit \cite{Herzog:2010hf} and the Fermi gas formalism \cite{Marino:2011eh}, all of which are based on the localization formula for the partition function obtained in \cite{Kapustin:2009kz}. 
Among them, the Fermi gas formalism has turned out to be quite powerful. 
It allows us to obtain, for example, the free energy, not only to all orders in $1/N$ expansion \cite{Fuji:2011km}, but even including non-perturbative terms \cite{Hatsuda:2012dt,Hatsuda:2013gj,Hatsuda:2013oxa}. 
Recently, such an analysis has been extended to Chern-Simons-matter theories whose dual theories contain orientifolds \cite{Moriyama:2015asx,Okuyama:2015auc,Honda:2015rbb,Okuyama:2016xke,Moriyama:2016xin,Moriyama:2016kqi}. 

In this paper, on the other hand, we revisit the analysis of planar solutions based on the matrix model technique. 
Although the reach of this technique is practically confined in the leading order of $1/N$ expansion unless the matrix model under consideration is simple enough, it can be applied to much wider family of Chern-Simons-matter theories, compared to the other methods. 
The aim of our research is to investigate a pattern in the strong coupling behaviors of various Chern-Simons-matter theories so that one could find a clue to know which theory could have a possible gravity dual. 
We expect that this line of research would shed some light on the underlying principle of how the space-time of the bulk gravity theory emerges from a quantum field theory. 

We focus our attention on a family of ${\cal N}\ge3$ Chern-Simons-matter theories with the gauge group ${\rm U}(N_1)\times {\rm U}(N_2)$ coupled to an arbitrary number $n$ of bi-fundamental hypermultiplets. 
The planar resolvents for such theories have been investigated in \cite{Suyama:2009pd}\cite{Marino:2009jd}\cite{Suyama:2010hr,Suyama:2011yz,Suyama:2013fua}, however, explicit expressions for the resolvents have not been obtained so far except for ABJM theory and ABJ theory \cite{Aharony:2008gk}. 
In this paper we show that, instead of the planar resolvent itself, its derivative can be determined explicitly for all theories mentioned above. 
More precisely, we define for each theory two planar resolvents which contain the information on two sets of eigenvalues in the matrix model. 
We determine the derivatives of both two resolvents explicitly except for the case $n=2$. 
For this exceptional case, which turns out to be the most interesting in the context of AdS/CFT correspondence, the derivative of a linear combination of the two resolvents is determined. 
From these results, we derive the explicit expressions of (a linear combination of) the vevs of BPS Wilson loops \cite{Drukker:2008zx,Chen:2008bp,Rey:2008bh} of the theories. 
Since the vevs can be written in terms of well-known functions, it is now straightforward to examine in which limit the vevs of the Wilson loops may diverge, the result of which provides an important hint for when a weak gravity dual might exist. 

This paper is organized as follows. 
In section \ref{pureCS}, we revisit the analysis of pure Chern-Simons theory in order to motivate us to consider the derivative of the planar resolvent. 
In section \ref{2-node}, we investigate the Chern-Simons-matter theories specified above, and determine explicitly the derivatives of the planar resolvents and the vevs of the Wilson loops. 
The analysis is done for the case $n=2$ (subsection \ref{n=2}) and for the other cases (subsection \ref{general-n}) separately. 
The validity of our formula for the planar resolvents is checked in section \ref{check} by calculating the vevs of the BPS Wilson loops perturbatively. 
Section \ref{discuss} is devoted to discussion. 

\vspace{1cm}

\section{Chern-Simons matrix model} \label{pureCS}

\vspace{5mm}

The partition function of the Chern-Simons matrix model is \cite{Marino:2002fk} 
\begin{equation}
Z\ =\ \int d^Nu\,\exp\left[ \frac{ik}{4\pi}\sum_{i=1}^N(u_i)^2 \right]\prod_{i<j}^N\sinh^2\frac{u_i-u_j}2. 
   \label{Z-pureCS}
\end{equation}
The overall constant which is irrelevant in the planar limit has been omitted. 
In the planar limit, any relevant quantities of this model are determined by the solution of the saddle-point equations 
\begin{equation}
\frac{k}{2\pi i}u_i\ =\ \sum_{j\ne i}\coth\frac{u_i-u_j}2. 
   \label{SPE-1}
\end{equation}
For example, the vev of the Wilson loop is given in terms of the solution as \cite{Kapustin:2009kz}
\begin{equation}
\langle W\rangle\ =\ \frac1N\sum_{i=1}^Ne^{u_i}. 
\end{equation}
The symmetry of the equations (\ref{SPE-1})  implies that the distribution of $\{u_i\}$ is invariant under the reflection, that is, the equality 
\begin{equation}
\{\, u_1, \cdots, u_N\ \}\ =\ \{\, -u_1, \cdots, -u_N\ \}
   \label{reflection}
\end{equation}
between two sets holds. 

It is convenient to introduce new variables $x_i:=e^{u_i}$ in terms of which (\ref{SPE-1}) can be written as 
\begin{equation}
\log x_i\ =\ \frac tN\sum_{j\ne i}\frac{x_i+x_j}{x_i-x_j}, 
   \label{SPE-x}
\end{equation}
where $t$ is the 't~Hooft coupling defined as 
\begin{equation}
t\ :=\ \frac{2\pi iN}{k}. 
\end{equation}
The condition (\ref{reflection}) for $u_i$ is translated to 
\begin{equation}
\{\ x_1, \cdots, x_N\ \}\ =\ \{\ x_1^{-1}, \cdots, x_N^{-1}\ \}. 
   \label{inversion}
\end{equation}

The solution of (\ref{SPE-x}) can be encoded in the resolvent defined as 
\begin{equation}
v(z)\ :=\ \frac tN\sum_{i=1}^N\frac{z+x_i}{z-x_i}. 
   \label{resolvent-pureCS}
\end{equation}
The large $z$ expansion 
\begin{equation}
v(z)\ =\ t+2t\langle W\rangle z^{-1}+O(z^{-2}) 
\end{equation}
provides us with the interesting physical quantities. 

Suppose that $t>0$. 
Then the equations (\ref{SPE-x}) can be interpreted as the equations for $N$ particles, lying on the real axis in $\mathbb{C}$, interacting among them and with an external log-type force. 
In this system, all $x_i>0$ are distributed around $x=1$. 
In the planar limit, the distribution of the eigenvalues $x_i$ becomes dense, and form an interval $[a,b]$ with $0<a<b$. 
The equality (\ref{inversion}) implies $ab=1$. 
The resolvent $v(z)$ becomes a holomorphic function on $\mathbb{C}\backslash [a,b]$ with a branch cut on $[a,b]$. 

As $t$ changes continuously to a complex value, the branch points $z=a,b$ move around in $\mathbb{C}$ while keeping $ab=1$ satisfied. 
We denote the branch cut by $[a,b]$ even when it does not lie on the real axis. 

The equations (\ref{SPE-x}) in the planar limit can be written in terms of $v(z)$ as 
\begin{equation}
2\log x\ =\ v(x_+)+v(x_-), \hspace{1cm} x\in[a,b]
   \label{saddle-v(z)}
\end{equation}
where $x_+$ ($x_-$) is the point in $\mathbb{C}$ slightly above (below) $x$ on the branch cut $[a,b]$. 
Requiring the finiteness of $v(z)$ at the branch points and at infinity, $v(z)$ is uniquely determined as 
\begin{equation}
v(z)\ =\ 2\log\frac{z+1-\sqrt{(z-a)(z-b)}}{\sqrt{a}+\sqrt{b}}. 
   \label{v(z)-pureCS}
\end{equation}
Note that the square-root is defined such that $\sqrt{(z-a)(z-b)}\to z$ for large positive $z$. 
The definition (\ref{resolvent-pureCS}) implies $t=v(\infty)$. 
This relates $t$ with $a$ as 
\begin{equation}
t\ =\ 2\log\frac{\sqrt{a}+\sqrt{b}}2. 
\end{equation}

\vspace{5mm}

The expression (\ref{v(z)-pureCS}) looks rather complicated compared to the resolvents of Hermitian matrix models. 
The logarithmic form seems to suggest that the fundamental object of the Chern-Simons matrix model would not be $v(z)$ itself but the exponential of $v(z)$. 
Indeed, the spectral curve of this model is given as 
\begin{equation}
e^{v+t}-(z+1)e^{\frac12(v+t)}+e^tz\ =\ 0, 
   \label{curve}
\end{equation}
which plays a role in a relation between Chern-Simons theory and a topological string theory \cite{Aganagic:2002wv}. 

It would be desirable if $e^{v(z)}$ could be determined directly from the saddle-point equation (\ref{saddle-v(z)}). 
In fact, this can be realized for the Chern-Simons matrix model and the lens space matrix models \cite{Halmagyi:2003ze}. 
However, a generalization of the techniques used in \cite{Halmagyi:2003ze} suitable for other matrix models does not look straightforward. 

\vspace{5mm}

For the case of Hermitian matrix models, the spectral curve can be derived from the loop equation. 
See e.g. \cite{Itoyama:2009sc} for a recent application of the loop equation to Hermitian matrix models. 
It is interesting if the loop equation for the Chern-Simons matrix model would reproduce (\ref{curve}). 
For Hermitian matrix models, the loop equation is nothing but the Schwinger-Dyson equations which imply  that the partition function of the matrix model satisfies the Virasoro constraints \cite{Itoyama:1990mf,David:1990ge,Mironov:1990im}. 

The Virasoro constraints for the Chern-Simons matrix model were studied in \cite{Nedelin:2015mio}. 
According to \cite{Nedelin:2015mio}, the corresponding Schwinger-Dyson equations can be organized into 
\begin{equation}
\int d^Nu\sum_{m=1}^N\frac{\partial}{\partial u_m}\left( \frac1{z-e^{u_m}}\exp\left[ \frac{ik}{4\pi}\sum_{i=1}^N(u_i)^2 \right]\prod_{i<j}^N\sinh^2\frac{u_i-u_j}2 \right)\ =\ 0. 
\end{equation}
In the planar limit, this can be rewritten as 
\begin{equation}
z\,\omega(z)^2-t\,\omega(z)\ =\ \log z\cdot\omega(z)-g(z), \hspace{1cm} g(z)\ :=\ \frac tN\sum_{i=1}^N\frac{\log z-u_i}{z-e^{u_i}}, 
   \label{loop-1}
\end{equation}
where $\omega(z)$ is defined in terms of the solution of (\ref{SPE-1}) as 
\begin{equation}
\omega(z)\ :=\ \frac tN\sum_{i=1}^N\frac1{z-e^{u_i}}. 
\end{equation}
The function $g(z)$ turns out not to be a simple function like a polynomial, contrary to the case of Hermitian matrix models. 
The function $g(z)$ is free from the square-root branch cut, but instead, it has a logarithmic branch cut. 
The discontinuity along the cut is 
\begin{equation}
g(x_+)-g(x_-)\ =\ 2\pi i\,\omega(x), \hspace{1cm} x\in(-\infty,0].
\end{equation}
This implies that $g(z)$ can be given in terms of $\omega(z)$ as 
\begin{equation}
g(z)\ =\ \int_{-\infty}^0dx\,\frac{\omega(x)}{x-z}. 
\end{equation}
Therefore, the loop equation for the Chern-Simons matrix model is not an algebraic equation but the following non-linear integral equation
\begin{equation}
z\,\omega(z)^2-t\,\omega(z)\ =\ \log z\cdot\omega(z)-\int_{-\infty}^0dx\,\frac{\omega(x)}{x-z}. 
   \label{loop-integral}
\end{equation}
Unfortunately, this equation looks quite difficult to solve. 

It is interesting to notice that at least one can guess the analytic structure of $\omega(z)$ from the integral equation (\ref{loop-integral}). 
One may find that each of two terms in the right-hand side of the equation (\ref{loop-integral}) has a logarithmic branch cut, but they cancel exactly between them. 
The non-linear structure of the left-hand side suggests the existence of a square-root branch cut in $\omega(z)$. 
Let $x\in(-\infty,0]$ be a point in $\mathbb{C}$ and $\tilde{x}$ be the corresponding point on the second Riemann sheet of $\omega(z)$. 
Then $\omega(\tilde{x})$ is different from $\omega(x)$ appearing in the integral. 
As a result, on the second Riemann sheet, the cancellation in the right-hand side is incomplete, and a logarithmic branch cut appears in $\omega(z)$. 
This is indeed the expected analytic structure of $\omega(z)$ since it is related to $v(z)$ given by (\ref{v(z)-pureCS}) as 
\begin{equation}
v(z)\ =\ 2tz\,\omega(z)-t. 
\end{equation}
It would be very interesting to find how to solve the integral equation (\ref{loop-integral}) and its generalizations derived from various Chern-Simons-matter matrix models. 

\vspace{5mm}

We have observed that the logarithmic form of $v(z)$ makes the analysis of the Chern-Simons matrix model complicated. 
It is interesting to notice that, in addition to exponentiating $v(z)$, there is another way to avoid dealing with the logarithmic form of $v(z)$. 
If one takes the derivative of $v(z)$, one obtains 
\begin{equation}
zv'(z)\ =\ 1-\frac{z-1}{\sqrt{(z-a)(z-b)}}. 
\end{equation}
The large $z$ expansion of $zv'(z)$ is 
\begin{equation}
zv'(z)\ =\ -2t\langle W\rangle z^{-1}+O(z^{-2}), 
\end{equation}
which preserves the information on $\langle W\rangle$ and all the higher moments. 
The missing information on $v(\infty)=t$ can be recovered via 
\begin{equation}
t\ =\ -\frac12\int_C\frac{dz}{2\pi i}\frac{\log z}{z} zv'(z), 
\end{equation}
where $C$ is a contour encircling the branch cut $[a,b]$ counterclockwise and excluding the origin. 
Therefore, it turns out to be sufficient to determine $zv'(z)$ for the investigation of the Chern-Simons matrix model in the planar limit. 
One finds that $zv'(z)$ is a solution of the equation 
\begin{equation}
2\ =\ \omega(x_+)+\omega(x_-), \hspace{1cm} x\in[a,b], 
\end{equation}
which is obtained from the derivative of (\ref{saddle-v(z)}). 
The solution $\omega(z)$ is uniquely determined by requiring that it has the following properties: 
\begin{itemize}
\item $\omega(z)$ is a holomorphic function on $\mathbb{C}\backslash[a,b]$, 
\item $\sqrt{(z-a)(z-b)}\,\omega(z)$ is finite at $z=a$ and $z=b$, 
\item $\omega(z)=O(z^{-1})$ for large $z$, and 
\item $\omega(z)$ satisfies 
\begin{equation}
\omega(z^{-1})\ =\ \omega(z). 
\end{equation}
\end{itemize}
Note that the last condition is a consequence of (\ref{inversion}). 

\vspace{5mm}

In the next section, we determine (a linear combination of) the derivatives of the resolvents for a family of Chern-Simons-matter matrix models whose gauge group is of the form ${\rm U}(N_1)\times{\rm U}(N_2)$. 
We find that the resolvents and (a linear combination of) the vevs of the BPS Wilson loops can be written explicitly for all such matrix models. 

\vspace{1cm}

\section{Chern-Simons-matter matrix models with 2 nodes} \label{2-node}

\vspace{5mm}

In this section, we investigate a Chern-Simons-matter matrix model obtained via the supersymmetric localization from a Chern-Simons-matter theory with 
\begin{itemize}
\item ${\cal N}\ge3$ supersymmetry, 
\item the gauge group ${\rm U}(N_1)_{k_1}\times{\rm U}(N_2)_{k_2}$, and 
\item $n$ bi-fundamental hypermultiplets. 
\end{itemize}
The family of such theories includes ABJM theory, ABJ theory, GT theory \cite{Gaiotto:2009mv} and theories discussed in \cite{Suyama:2013fua}. 

\vspace{5mm}

The partition function of the matrix model is given as \cite{Kapustin:2009kz}
\begin{equation}
Z\ =\ \int d^{N_1}u\,d^{N_2}w\,\exp\left[ \frac{ik_1}{4\pi}\sum_{i=1}^{N_1}(u_i)^2+\frac{ik_2}{4\pi}\sum_{a=1}^{N_2}(w_a)^2 \right]\frac{\prod_{i<j}^{N_1}\sinh^2\frac{u_i-u_j}2\prod_{a<b}^{N_2}\sinh^2\frac{w_a-w_b}2}{\prod_{i=1}^{N_1}\prod_{a=1}^{N_2}\cosh^n\frac{u_i-w_a}2}. 
   \label{Z-2node}
\end{equation}
The saddle-point equations are 
\begin{eqnarray}
\frac{k_1}{2\pi i}u_i &=& \sum_{j\ne i}^{N_1}\coth\frac{u_i-u_j}2-\frac n2\sum_{a=1}^{N_2}\tanh\frac{u_i-w_a}2, \\
\frac{k_2}{2\pi i}w_a &=& \sum_{b\ne a}^{N_2}\coth\frac{w_a-w_b}2-\frac n2\sum_{i=1}^{N_1}\tanh\frac{w_a-u_i}2. 
\end{eqnarray}
In terms of new variables $x_i:=e^{u_i}$ and $y_a:=-e^{w_a}$, these equations can be written as 
\begin{eqnarray}
\frac{k_1}{2\pi i}\log x_i &=& \sum_{j\ne i}^{N_1}\frac{x_i+x_j}{x_i-x_j}-\frac n2\sum_{a=1}^{N_2}\frac{x_i+y_a}{x_i-y_a}, 
   \label{2N-saddle1} \\
\frac{k_2}{2\pi i}\log(-y_a) &=& \sum_{b\ne a}^{N_2}\frac{y_a+y_b}{y_a-y_b}-\frac n2\sum_{i=1}^{N_1}\frac{y_a+x_i}{y_a-x_i}. 
   \label{2N-saddle2}
\end{eqnarray}

To define the planar limit in a symmetric manner, we introduce an auxiliary parameter $k$ and define 
\begin{equation}
t_1\ :=\ \frac{2\pi iN_1}{k}, \hspace{1cm} t_2\ :=\ \frac{2\pi iN_2}{k}, \hspace{1cm} \kappa_1\ :=\ \frac{k_1}k, \hspace{1cm} \kappa_2\ :=\ \frac{k_2}k. 
   \label{parameters-2node}
\end{equation}
The planar limit is then defined as the limit $k\to\infty$ while keeping these parameters fixed. 
 
We define two resolvents for two sets $\{x_i\}, \{y_a\}$ of eigenvalues as 
\begin{equation}
v_1(z)\ :=\ \frac{t_1}{N_1}\sum_{i=1}^{N_1}\frac{z+x_i}{z-x_i}, \hspace{1cm} v_2(z)\ :=\ \frac{t_2}{N_2}\sum_{a=1}^{N_2}\frac{z+y_a}{z-y_a}. 
   \label{def-2Nres}
\end{equation}
In the planar limit, $v_1(z)$ becomes a holomorphic function on $\mathbb{C}\backslash[a_1,b_1]$, and $v_2(z)$ becomes a holomorphic function on $\mathbb{C}\backslash[a_2,b_2]$. 
As in the previous section, $a_1b_1=a_2b_2=1$ is assumed. 
In terms of these resolvents, the saddle-point equations (\ref{2N-saddle1})(\ref{2N-saddle2}) can be written as 
\begin{eqnarray}
2\kappa_1\log x &=& v_1(x_+)+v_1(x_-)-n\,v_2(x), \hspace{1cm} x\in[a_1,b_1], \\
2\kappa_2\log(-y) &=& v_2(y_+)+v_2(y_-)-n\,v_1(y), \hspace{1.1cm} y\in[a_2,b_2].
\end{eqnarray}

Our observation in the previous section suggests that, instead of dealing with these equations, we should investigate the following equations 
\begin{eqnarray}
{2\kappa_1} &=& xv_1'(x_+)+xv_1'(x_-)-n\,xv_2'(x), 
   \label{res-eq1} \\
{2\kappa_2} &=& yv_2'(y_+)+yv_2'(y_-)-n\,yv_1'(y). 
   \label{res-eq2}
\end{eqnarray}
It is convenient to combine the two resolvents into a vector-valued resolvent 
\begin{equation}
v(z)\ :=\ (v_1(z), v_2(z)). 
\end{equation}
In terms of $v(z)$, the equations (\ref{res-eq1})(\ref{res-eq2}), together with 
\begin{equation}
v_1(y_+)\ =\ v_1(y_-), \hspace{1cm} v_2(x_+)\ =\ v_1(x_-), 
\end{equation}
which are required by the definition (\ref{def-2Nres}), can be written as follows: 
\begin{eqnarray}
\left( {2\kappa_1}, 0 \right) &=& xv'(x_+)-xv'(x_-)M_1, 
   \label{res-vec1} \\
\left( 0, {2\kappa_2} \right) &=& yv'(y_+)-yv'(y_-)M_2, 
   \label{res-vec2}
\end{eqnarray}
where 
\begin{equation}
M_1\ :=\ \left[
\begin{array}{cc}
-1 & 0 \\
n & 1
\end{array}
\right], \hspace{1cm} M_2\ :=\ \left[
\begin{array}{cc}
1 & n \\
0 & -1
\end{array}
\right]. 
\end{equation}
The properties required for the solution of (\ref{res-vec1})(\ref{res-vec2}) are as follows: 
\begin{itemize}
\item $zv'(z)$ is holomorphic on $\mathbb{C}\backslash([a_1,b_1]\cup[a_2,b_2])$, 
\item $s(z)zv'(z)$ is finite at the branch points, where 
\begin{equation}
s(z)\ :=\ \sqrt{(z-a_1)(z-b_1)(z-a_2)(z-b_2)}, 
   \label{s(z)}
\end{equation}
\item $zv'(z)=O(z^{-1})$ for large $z$, and 
\item $zv'(z)$ satisfies 
\begin{equation}
z^{-1}v'(z^{-1})\ =\ zv'(z). 
   \label{inversion2}
\end{equation}
\end{itemize}

The 't Hooft couplings $t:=(t_1,t_2)$ are given as 
\begin{equation}
(t_1,0)\ =\ -\frac12\int_{C_1}\frac{dz}{2\pi i}\frac{\log z}{z} zv'(z), \hspace{1cm} (0,t_2)\ =\ -\frac12\int_{C_2}\frac{dz}{2\pi i}\frac{\log z}{z} zv'(z), 
\end{equation}
where $C_1$ and $C_2$ are contours encircling $[a_1,b_1]$ and $[a_2,b_2]$ counterclockwise, respectively, and excluding the origin. 
The vevs of the BPS Wilson loops are obtained from the large $z$ expansion of $zv'(z)$ as 
\begin{equation}
zv'(z)\ =\ -2\left( t_1\langle W_1\rangle, -t_2\langle W_2\rangle \right)z^{-1}+O(z^{-2}). 
   \label{WL-general}
\end{equation}
Note that there is a minus sign in front of $\langle W_2\rangle$ since 
\begin{equation}
\langle W_2\rangle\ =\ \frac1{N_2}\sum_{a=1}^{N_2}e^{w_a}\ =\ -\frac1{N_2}\sum_{a=1}^{N_2}y_a. 
\end{equation}

\vspace{5mm}

\subsection{The case $n=2$} \label{n=2}

\vspace{5mm}

First, we consider the case $n=2$. 
The matrix model with $n=2$ corresponds to ABJM theory and ABJ theory when $\kappa_1+\kappa_2=0$. 
In general ($\kappa_1+\kappa_2\ne0$), the matrix model is derived from GT theory which is expected to describe a massive Type IIA theory. 

In the case $n=2$, the equations (\ref{res-vec1})(\ref{res-vec2}) can be simplified as follows. 
Notice that the matrices $M_1$ and $M_2$ have a common eigenvector: 
\begin{eqnarray}
\left[
\begin{array}{cc}
-1 & 0 \\
2 & 1
\end{array}
\right]\left[
\begin{array}{c}
1 \\
-1
\end{array}
\right] 
\ =\  -\left[
\begin{array}{c}
1 \\
-1 
\end{array}
\right] \ =\ 
\left[
\begin{array}{cc}
1 & 2 \\
0 & -1
\end{array}
\right]\left[
\begin{array}{c}
1 \\
-1
\end{array}
\right]. 
\end{eqnarray}
Multiplying this eigenvector from the right, the equations (\ref{res-vec1})(\ref{res-vec2}) become 
\begin{eqnarray}
{2\kappa_1} &=& \omega(x_+)+\omega(x_-), 
   \label{res1} \\
-{2\kappa_2} &=& \omega(y_+)+\omega(y_-), 
   \label{res2}
\end{eqnarray}
where $\omega(z)$ is defined as 
\begin{equation}
\omega(z)\ :=\ zv'(z)\left[
\begin{array}{c}
1 \\
-1
\end{array}
\right]\ =\ zv_1'(z)-zv_2'(z). 
\end{equation}
The required properties for $zv'(z)$ is translated to those of $\omega(z)$ by this definition. 
Once $\omega(z)$ is determined, one can show that $zv'(z)$ can be given in terms of integrals of $\omega(z)$. 

The function $\omega(z)$ already contains a lot of information. 
For example, the 't Hooft couplings are given as 
\begin{equation}
t_1\ =\ -\frac12\int_{C_1}\frac{d\xi}{2\pi i}\frac{\log z}z\omega(z), \hspace{1cm} t_2\ =\ \frac12\int_{C_2}\frac{d\xi}{2\pi i}\frac{\log z}z\omega(z), 
\end{equation}
and the large $z$ expansion of $\omega(z)$ gives 
\begin{equation}
\omega(z)\ =\ -2(t_1\langle W_1\rangle+t_2\langle W_2\rangle)z^{-1}+O(z^{-2}). 
\end{equation}
Note that the linear combination of $\langle W_1\rangle$ and $\langle W_2\rangle$ appearing above gives the vev of the half-BPS Wilson loop \cite{Drukker:2009hy}, in the case of ABJM theory and ABJ theory. 

\vspace{5mm}

The solution of (\ref{res1})(\ref{res2}) with the required properties can be given as follows. 
Let $\Omega(z,\xi)$ be a holomorphic function on $\mathbb{C}\backslash\{a_1,b_1,a_2,b_2,\xi\}$ with a parameter $\xi$, satisfying the following conditions: 
\begin{itemize}
\item $\Omega(z,\xi)$ has a monodromy $-1$ at the points $z=a_1,b_1,a_2,b_2$, and 
\item $\Omega(z,\xi)$ has a simple pole at $z=\xi$ with the residue $1$. 
\end{itemize}
Using these properties of $\Omega(z,\xi)$, One can easily check that 
\begin{equation}
\omega_0(z)\ :=\ \kappa_1\int_{C_1}\frac{d\xi}{2\pi i}\Omega(z,\xi)-\kappa_2\int_{C_2}\frac{d\xi}{2\pi i}\Omega(z,\xi) 
\end{equation}
is a solution of (\ref{res1})(\ref{res2}). 
The finiteness of $s(z)\omega_0(z)$ at the branch points suggests that an appropriate choice of $\Omega(z,\xi)$ is 
\begin{equation}
\Omega(z,\xi)\ =\ \frac{h(z)}{h(\xi)}\frac1{z-\xi}\frac{s(\xi)}{s(z)}
\end{equation}
with $h(z)$ an entire function. 
A convenient choice turns out to be $h(z)=z$. 

One finds that $\omega_0(z)$ does not satisfy the inversion condition 
\begin{equation}
\omega(z^{-1})\ =\ \omega(z) 
   \label{inversion3}
\end{equation}
deduced from (\ref{inversion2}).  
This problem is remedied by noticing that $\omega_0(z^{-1})$ also satisfies the equations (\ref{res1})(\ref{res2}). 
Therefore, 
\begin{equation}
\omega(z)\ =\ \frac12\omega_0(z)+\frac1{2}\omega_0(z^{-1})
\end{equation}
is a solution which also satisfies the inversion condition (\ref{inversion3}). 
One can check that this is the only solution of (\ref{res1})(\ref{res2}) which has all the required properties deduced from those of $zv'(z)$. 

By deforming the integration contour, $\omega(z)$ can be written as 
\begin{equation}
\omega(z)\ =\ -{\kappa_2}\left[ 1-\frac{z^2-1}{s(z)} \right]-\frac{\kappa_1+\kappa_2}2\frac{z^2-1}{s(z)}F(z). 
\end{equation}
The function $F(z)$ defined as 
\begin{equation}
F(z)\ :=\ \int_{C_1}\frac{d\xi}{2\pi i}\frac{s(\xi)}{\xi(\xi-z)(\xi-z^{-1})}, 
\end{equation}
which is absent for ABJM theory and ABJ theory ($\kappa_1+\kappa_2=0$), can be written in terms of the complete elliptic integrals. 
Explicitly, 
\begin{eqnarray}
F(z) 
&=& \frac{8\alpha}{\sqrt{(1-\alpha^2)(1-k^2\alpha^2)}}\frac{f(\alpha^{-2})-f(k(z)^2)}{(2-z-z^{-1})(1-k(z)^2\alpha^2)}, 
\end{eqnarray}
where 
\begin{equation}
f(z)\ :=\ -\frac{2(1-z)(k^2-z)}{\pi i z}\Pi_1(-z,k)-\frac2{\pi i}E(k)+\frac{2(1-z)k^2}{\pi iz}K(k), 
\end{equation}
and 
\begin{equation}
\alpha\ :=\ \frac{1+a_1}{1-a_1}, \hspace{1cm} k(z)^2\ := \frac{z+z^{-1}+2}{z+z^{-1}-2}\alpha^{-2}, \hspace{1cm} k^2\ :=\ k(a_2)^2. 
\end{equation}

The planar resolvent for ABJM theory and ABJ theory was obtained in \cite{Marino:2009jd} using the result of \cite{Halmagyi:2003ze}. 
The resolvent $\omega(z)$ determined above for the case $\kappa_1+\kappa_2=0$ can be derived from the result of \cite{Marino:2009jd}. 
We have found that the resolvent for general $\kappa_1$ and $\kappa_2$ has a quite complicated expression  compared to that for the case $\kappa_1+\kappa_2=0$. 

\vspace{5mm}

The large $z$ expansion of $\omega(z)$ gives 
\begin{eqnarray}
& & t_1\langle W_1\rangle+t_2\langle W_2\rangle \nonumber \\
&=& -\frac{\kappa_2}{4}(a_1+\cdots+b_2)+\frac{2\alpha(\kappa_1+\kappa_2)}{\pi i\sqrt{(1-\alpha^2)(1-k^2\alpha^2)}}\left[ \frac{1-k^2\alpha^4}{\alpha^2}\Pi_1(-\alpha^{-2},k)-E(k)+(1+k^2\alpha^2)K(k) \right]. \nonumber \\
   \label{n=2 WL}
\end{eqnarray}

Recall that, in various examples of AdS/CFT correspondence, the vev of a Wilson loop diverges as 
\begin{equation}
\log |\langle W\rangle|\ =\ c\lambda^\gamma, 
\end{equation}
in the limit where the 't Hooft coupling $\lambda$ is large. 
For example, the exponent $\gamma$ is $\frac12$ for ${\cal N}=4$ super Yang-Mills theory \cite{Erickson:2000af}\cite{Pestun:2007rz} and ABJM theory \cite{Suyama:2009pd}\cite{Marino:2009jd}, and $\gamma=\frac13$ for GT theory \cite{Suyama:2011yz}. 
Therefore, one may be interested in a divergent behavior of the expression (\ref{n=2 WL}) since it would be a sign of a possible existence of a dual gravity description via AdS/CFT correspondence. 
Obviously, the first term of (\ref{n=2 WL}) diverges when $b_1$ or $b_2$ diverges, or in other words, when  $a_1\to0$ or $a_2\to0$. 
The second term of (\ref{n=2 WL}) is divergent if $\alpha\to1$ or $k\to1$, which correspond to 
$a_1\to0$ or $a_2\to a_1$, respectively. 
The expression (\ref{n=2 WL}) shows that there is no other divergent behavior. 

It was observed, e.g. in \cite{Suyama:2011yz}, that the simultaneous limit $\alpha\to1$ and $k\to1$ corresponds to the limit in which a weak gravity dual exists. 
In another limit, say $\alpha\to1$ but $k$ is different from 1, the distribution of two sets of eigenvalues becomes hierarchical, that is, the distribution of $\{x_i\}$ becomes large while that of $\{y_a\}$ is not. 
In such a situation, the two sets of equations (\ref{2N-saddle1})(\ref{2N-saddle2}) would decouple effectively, and each set of equations would become similar to the saddle-point equations (\ref{SPE-x}) for the Chern-Simons matrix model. 

\vspace{5mm}

\subsection{The cases $n\ne 2$} \label{general-n}

\vspace{5mm}

Next, consider the other cases $n\ne2$. 
Recall that we would like to solve the following equations 
\begin{eqnarray}
\left( {2\kappa_1}, 0 \right) &=& xv'(x_+)-xv'(x_-)M_1, \\
\left( 0, {2\kappa_2} \right) &=& yv'(y_+)-yv'(y_-)M_2. 
\end{eqnarray}
For the cases $n\ne2$, a constant vector $c:=(c_1,c_2)$ satisfies these equations since there exist the constants $c_1, c_2$ which satisfy 
\begin{equation}
(2\kappa_1, 2\kappa_2)\ =\ (c_1, c_2)\left[
\begin{array}{cc}
2 & -n \\
-n & 2
\end{array}
\right]. 
\end{equation}

Define a function $\omega(z)$ such that $zv'(z)$ is given as 
\begin{equation}
zv'(z)\ =\ c+\omega(z). 
\end{equation}
Then, $\omega(z)$ satisfies 
\begin{equation}
\omega(x_+) \ =\  \omega(x_-)M_1, \hspace{1cm} 
\omega(y_+) \ =\  \omega(y_-)M_2.  
   \label{eq-omega}
\end{equation}
It is convenient to consider, instead of $\omega(z)$, a function $f(z)$ defined as 
\begin{equation}
f(z)\ :=\ s(z)\omega(z)
\end{equation}
which is required to have the following properties: 
\begin{itemize}
\item $f(z)$ is holomorphic on $\mathbb{C}\backslash([a_1,b_1]\cup[a_2,b_2])$, 
\item $f(z)$ is finite at the branch points, 
\item for large $z$, $f(z)$ behaves as 
\begin{equation}
f(z)\ =\ -cz^2+O(z), 
   \label{f-infty}
\end{equation}
\item $f(z)$ satisfies 
\begin{equation}
f(z^{-1})\ =\ -z^{-2}f(z). 
\end{equation}
\end{itemize}
The equations (\ref{eq-omega}) can be written in terms of $f(z)$ as 
\begin{equation}
f(x_+) \ =\  -f(x_-)M_1, \hspace{1cm} 
f(y_+) \ =\  -f(y_-)M_2.  
   \label{eq-f}
\end{equation}
We will show that the solution of (\ref{eq-f}) with the above properties is uniquely determined. 

\vspace{5mm}

The problem of determining $f(z)$ turns out to be a generalization of the problem in  \cite{Eynard:1995nv}\cite{Eynard:1995zv} discussing the ${\rm O}(n)$ model \cite{Kostov:1988fy}\cite{Gaudin:1989vx}, and therefore, the analysis developed in \cite{Eynard:1995nv}\cite{Eynard:1995zv} can be applied to our problem with a suitable modification. 

Our strategy is to map the double cover of $\mathbb{C}\backslash([a_1,b_1]\cup[a_2,b_2])$ to a torus $T^2$ by a map defined as 
\begin{equation}
u(z)\ :=\ \frac{\varphi(z)}{2\varphi(b_1)}, \hspace{1cm} \varphi(z)\ :=\ \int_{a_1}^z\frac{d\xi}{s(\xi)}, 
\end{equation}
where the integration contour for $\varphi(b_1)$ lies above the branch cut $[a_1,b_1]$. 
Note that $u(z)$ satisfies 
\begin{equation}
u(z^{-1})\ =\ u(z)-\frac12. 
\end{equation}
Let $\tau:=2u(a_2)$ be the modulus of $T^2$. 
The function $f(z)$ becomes a function on $T^2$ by the inverse map
\begin{equation}
z(u)\ :=\ -\frac{\vartheta_1(u-u_0)\vartheta_1(u+u_0)}{\vartheta_1(u-u_\infty)\vartheta_1(u+u_\infty)}, 
\end{equation}
where $\vartheta_1(u):=\vartheta_1(u,\tau)$ is the theta function, and $u_z:=u(z)$. 
In the following, the function $f(z(u))$ on $T^2$ is denoted simply by $f(u)$. 

By the definition of the $u$-coordinate, $f(u)$ satisfies 
\begin{equation}
f(u+1)\ =\ f(u), 
   \label{period}
\end{equation}
since the shift of $u$ by $1$ corresponds to a move around the branch cut $[a_1,b_1]$ in the $z$-plane. 
The equations (\ref{eq-f}) can be written as 
\begin{eqnarray}
f(-u) &=& -f(u)M_1, 
   \label{f1 to f2} \\
f(u+\tau) &=& f(u)M_1M_2. 
   \label{eq-f_2}
\end{eqnarray}
The matrix $M_1M_2$ can be diagonalized by a matrix $S$ defined as 
\begin{equation}
S\ := \ \left[
\begin{array}{cc}
1 & 1 \\
-e^{\pi i\nu} & -e^{-\pi i\nu}
\end{array}
\right], 
\end{equation}
where $\nu$ parametrizes $n$ as $n=2\cos\pi\nu$. 
Therefore, the equations (\ref{period})(\ref{eq-f_2}) for a vector-valued function $f(u)$ can be split into two sets of equations for two scalar-valued functions. 
Define $(\tilde{f}_1(u),\tilde{f}_2(u)):=f(u)S$. 
Then $\tilde{f}_1(u)$ satisfies 
\begin{equation}
\tilde{f}_1(u+1)\ =\ \tilde{f}_1(u), \hspace{1cm} \tilde{f}_1(u+\tau)\ =\ e^{2\pi i\nu}\tilde{f}_1(u). 
   \label{eq-tilde{f}}
\end{equation}
Note that $S$ also simplifies $M_1$ and $M_2$ separately as 
\begin{equation}
S^{-1}M_1S\ =\ \left[
\begin{array}{cc}
0 & -1 \\
-1 & 0
\end{array}
\right], \hspace{1cm} S^{-1}M_2S\ =\ \left[
\begin{array}{cc}
0 & -e^{-2\pi i\nu} \\
-e^{2\pi i\nu} & 0
\end{array}
\right]. 
\end{equation}
The equation (\ref{f1 to f2}) relates $\tilde{f}_2(u)$ to $\tilde{f}_1(u)$ as 
\begin{equation}
\tilde{f}_2(u)\ =\ \tilde{f}_1(-u). 
\end{equation}
In the following, we determine $\tilde{f}_1(u)$ which has the required properties deduced from those of $f(z)$. 

\vspace{5mm}

A solution $G(u)$ of the equations (\ref{eq-tilde{f}}) can be constructed in terms of the theta functions, although it is not uniquely determined. 
Our choice of $G(u)$ is 
\begin{equation}
G(u)\ :=\ \frac{\vartheta_1(u-u_\nu)\vartheta_1(u-u_\nu+\frac12)}{\vartheta_1(u-u_\infty)\vartheta_1(u+u_\infty)}, 
\end{equation}
where $u_\nu:=\frac12\nu+\frac14$. 
An advantage of this choice is that $G(u)$ has a nice inversion property  
\begin{equation}
G(u(z^{-1}))\ =\ -\frac1zG(u(z)). 
   \label{inversion-G}
\end{equation}
The product $g(u):=\tilde{f}_1(u)G(u)^{-1}$ then satisfies 
\begin{equation}
g(u+1)\ =\ g(u), \hspace{1cm} g(u+\tau)\ =\ g(u), 
\end{equation}
that is, $g(u)$ is an elliptic function. 

Since $\tilde{f}_1(z)$ has a double pole at infinity and otherwise finite, $g(u)$ must have simple poles at $u=u_\infty, -u_\infty, u_\nu, u_\nu-\frac12$. 
The Riemann-Roch theorem implies that elliptic functions with at most four such simple poles form a four-dimensional vector space $V$. 
Therefore, $g(u)$ can be written as 
\begin{equation}
g(u)\ =\ r_1g_1(u)+r_2g_2(u)+r_3g_3(u)+r_4g_4(u),
\end{equation}
when a basis of $V$ is given. 
We choose a basis as 
\begin{equation}
\begin{array}{rclcrcl}
g_1(u) & := & 1, & \hspace{5mm} & g_2(u) & := & \displaystyle{-\frac{\vartheta_1(u-u_0)\vartheta_1(u+u_0)}{\vartheta_1(u-u_\infty)\vartheta_1(u+u_\infty)}}, \\ [4mm]
g_3(u) & := & \displaystyle{\frac{\vartheta_1(u-u_0)\vartheta_1(u-u_\nu+\frac12)}{\vartheta_1(u-u_\infty)\vartheta_1(u-u_\nu)}}, & \hspace{5mm} & g_4(u) & := & \displaystyle{-\frac{\vartheta_1(u+u_0)\vartheta_1(u-u_\nu)}{\vartheta_1(u+u_\infty)\vartheta_1(u-u_\nu+\frac12)}}. 
\end{array}
   \label{basis}
\end{equation}

Due to the inversion property (\ref{inversion-G}) of $G(u)$, the elliptic function $g(u)$ is required to satisfy 
\begin{equation}
g(u(z^{-1}))\ =\ \frac1zg(u(z)). 
\end{equation}
This condition implies
\begin{equation}
r_1\ =\ r_2, \hspace{1cm} r_3\ =\ r_4. 
\end{equation}
The remaining two coefficients, say $r_1$ and $r_3$, are fixed by requiring the asymptotic behavior (\ref{f-infty}) of $f(z)$ at infinity. 
Equivalently, they are determined by requiring $f(0)=c$. 
In terms of the $u$ variable, this implies 
\begin{equation}
\tilde{f}_1(u_0)\ =\ \tilde{c}_1, \hspace{1cm} \tilde{f}_2(u_0)\ =\ \tilde{f}_1(-u_0)\ =\ \tilde{c}_2, 
\end{equation}
where $\tilde{c}:=cS$. 
The solution is 
\begin{eqnarray}
r_1 
&=& \frac1{g_3(-u_0)-g_4(u_0)}\left[ \tilde{c}_1\frac{g_3(-u_0)}{G(u_0)}-\tilde{c}_2\frac{g_4(u_0)}{G(-u_0)} \right], 
   \label{r_1} \\
r_3 
&=& \frac1{g_3(-u_0)-g_4(u_0)}\left[ -\frac{\tilde{c}_1}{G(u_0)}+\frac{\tilde{c}_2}{G(-u_0)} \right]. 
   \label{r_3}
\end{eqnarray}

Now, the elliptic function $g(u)$ has been determined completely. 
The resolvent $zv'(z)$ is therefore given in terms of the theta functions explicitly as 
\begin{equation}
zv'(z)\ =\ c+\left[ \frac1{s(z)}g(u(z))G(u(z)),\ \frac1{s(z)}g(-u(z))G(-u(z))\, \right]S^{-1}. 
   \label{resolvent-general}
\end{equation}

Note that a part of the above calculations can be applied to the case $n=2$ as long as  $\kappa_1+\kappa_2=0$. 
In fact, the multiplication by $S$ for the case $n=2$ corresponds to the multiplication by ${1\brack-1}$ used in subsection \ref{n=2}. 

\vspace{5mm}

One can show that the 't~Hooft couplings can be written as 
\begin{equation}
t_1\ =\ -\frac12\int_{C_1}\frac{dz}{2\pi i}\frac{\log z}{z}\frac{\tilde{f}_1(z)}{s(z)}, \hspace{1cm} t_2\ =\ \frac12e^{-\pi i\nu}\int_{C_2}\frac{dz}{2\pi i}\frac{\log z}{z}\frac{\tilde{f}_1(z)}{s(z)}. 
\end{equation}

The large $z$ expansion (\ref{WL-general}) of the resolvent gives the vevs of the BPS Wilson loops. 
Equivalently, they can be obtained from the small $z$ expansion of $zv'(z)$: 
\begin{eqnarray}
zv'(z) 
&=& c-f(0)-\left[ \frac{a_1+\cdots+b_2}2f(0)+f'(0) \right]z+O(z^2). 
\end{eqnarray}
We imposed $f(0)=c$ to determine the elliptic function $g(u)$. 
Then, the vevs of the BPS Wilson loops are 
\begin{equation}
\left( t_1\langle W_1\rangle,-t_2\langle W_2\rangle \right)\ =\ \frac{a_1+\cdots+b_2}4c+\frac12f'(0). 
\end{equation}

\vspace{5mm}

Note that the coefficients $r_1$ and $r_3$ given in (\ref{r_1})(\ref{r_3}) may diverge for a particular configuration of the branch cuts. 
Recall that $u_0$ is a function of the positions $a_1,\cdots,b_2$. 
One can show that, as functions of $u_0$, $r_1$ and $r_3$ have poles at $u_0=u_\nu,-u_\nu+\frac12$ and at values such that $g_3(-u_0)=g_4(u_0)$. 

The former cases, it is easy to show that the basis functions (\ref{basis}) degenerate as 
\begin{equation}
g_3(u)\ =\ \left\{
\begin{array}{cc}
1, & (u_0=u_\nu), \\ [2mm]
-g_2(u), & (u_0=-u_\nu+\frac12)
\end{array}
\right. \hspace{1cm} 
g_4(u)\ =\ \left\{
\begin{array}{cc}
g_2(u), & (u_0=u_\nu), \\ [2mm]
-1. & (u_0=-u_\nu+\frac12)
\end{array}
\right. 
\end{equation}
Due to these degenerations, the poles are canceled among them, and therefore, $g(u)$ is finite for generic $u$. 
Since the 't~Hooft couplings and the vevs of the Wilson loops can be given in terms of contour integrals, the finiteness of $g(u)$ implies the finiteness of these quantities. 
Therefore, the poles at $u_0=u_\nu,-u_\nu+\frac12$ are physically irrelevant. 

The latter case corresponds to the case $z(u_\nu)=-1$. 
This implies 
\begin{equation}
u_\nu\ =\ \frac12\tau\pm\frac14 \mbox{ mod }\mathbb{Z}+\mathbb{Z}\tau. 
\end{equation}
In terms of $\nu$, this condition is written as 
\begin{equation}
\nu\ =\ \tau \mbox{ mod }\mathbb{Z}+2\mathbb{Z}\tau. 
\end{equation}
When $\tau$, which is also a function of the positions $a_1,\cdots,b_2$, is chosen such that the above equation holds for a given $\nu$, then $g(u)$ satisfies 
\begin{equation}
(g_3(-u_0)-g_4(u_0))g(u)\ =\ \left[ \frac{\tilde{c}_1}{G(u_0)}-\frac{\tilde{c}_2}{G(-u_0)} \right]\Bigl[ g_3(-u_0)(1+g_2(u))-g_3(u)-g_4(u) \Bigr]. 
\end{equation}
Since the basis functions (\ref{basis}) are linearly independent, the right-hand side is not identically zero. 
Therefore, $g(u)$ diverges for generic $u$ when $z(u_\nu)=-1$ holds. 

One can check that, for example when $a_1=-a_2$ holds, the quantities $t_1\langle W_1\rangle$ and $t_2\langle W_2\rangle$ diverge. 
Note that the definitions of $\langle W_1\rangle$ and $\langle W_2\rangle$ imply 
\begin{equation}
|\langle W_1\rangle|\ \le\ |b_1|, \hspace{1cm} |\langle W_2\rangle|\ \le\ |b_2|. 
\end{equation}
Unless the two branch cuts are hierarchical, a finite $\tau$ corresponds to finite $b_1$ and $b_2$, implying that the vevs $\langle W_1\rangle$ and $\langle W_2\rangle$ are finite. 
Therefore, the divergence for $z(u_\nu)=-1$ is due to the divergence of the 't~Hooft couplings $t_1$ and $t_2$. 
This means that there exists a large 't~Hooft coupling limit in the parameter space of a 2-node theory with $n>2$ at which the vevs of the Wilson loops are finite. 
A similar kind of behavior was observed in \cite{Suyama:2013fua} for more general theories. 

\vspace{1cm}

\section{Perturbative check} \label{check}

\vspace{5mm}

In this section, we will use the planar resolvent obtained in section \ref{2-node} for the  calculation of the vevs of the Wilson loops for the 2-node theories perturbatively. 
The same perturbative expansion can be also obtained directly from their localization formulas. 
The match between these two results provides a non-trivial check for the validity of our formulas for the planar resolvents. 

\vspace{5mm}

\subsection{Expansion from the localization formula}

\vspace{5mm}

The vev of a Wilson loop can be given in terms of a finite-dimensional integral via the supersymmetric localization \cite{Kapustin:2009kz}. 
For pure Chern-Simons theory, the vev $\langle W\rangle$ is given as 
\begin{equation}
\langle W\rangle\ =\ \frac1Z\int d^Nu\ \exp\left[ \frac{ik}{4\pi}\sum_{i=1}^N(u_i)^2 \right]\prod_{i<j}^N\sinh^2\frac{u_i-u_j}2\cdot \frac1N\sum_{i=1}^Ne^{u_i}, 
   \label{W-CS}
\end{equation}
where $Z$ is defined as (\ref{Z-pureCS}). 
The $1/k$ expansion of $\langle W\rangle$ can be derived in a manner explained in \cite{Kapustin:2009kz}. 
The idea is to relate the vev (\ref{W-CS}) to the vevs of the Gaussian matrix model whose partition function $Z_0$ is defined as 
\begin{eqnarray}
Z_0\ :=\ \int d^Nu\,\exp\left[ -\frac12\sum_{i=1}^N(u_i)^2 \right]\prod_{i=1}^N(u_i-u_j)^2. 
\end{eqnarray}

The partition function $Z$ can be rewritten as follows: 
\begin{eqnarray}
Z 
&=& 2^{-N(N-1)}\left( \frac{2\pi i}{k} \right)^{\frac12N^2}\int d^Nu\,\exp\left[ -\frac12\sum_{i=1}^N(u_i)^2 \right]\prod_{i<j}^N(u_i-u_j)^2\cdot \sum_{n=0}^\infty \left( \frac{2\pi i}k \right)^nX_n(u), 
\end{eqnarray}
where $X_n(u)$ are defined such that 
\begin{eqnarray}
\exp\left[ \sum_{i<j}^N2\log\left( \frac{\sinh\sqrt{\frac{2\pi i}k}\frac{u_i-u_j}2}{\sqrt{\frac{2\pi i}k}\frac{u_i-u_j}2} \right) \right] 
&=& \sum_{n=0}^\infty \left( \frac{2\pi i}k \right)^nX_n(u) \nonumber \\
&=& 1+\frac{2\pi i}k\cdot\frac13\sum_{i<j}^N\left( \frac{u_i-u_j}2 \right)^2+O(k^{-2}). 
\end{eqnarray}
The same rewriting can be also performed in the presence of an operator insertion. 
Therefore, the vev (\ref{W-CS}) can be written as 
\begin{eqnarray}
\langle W\rangle 
&=& \frac{\displaystyle{\left\langle \sum_{n=0}^\infty\left( \frac{2\pi i}k \right)^nX_n(u)\sum_{m=0}^\infty\left( \frac{2\pi i}k \right)^mW_m(u) \right\rangle_0}}{\displaystyle{\left\langle \sum_{n=0}^\infty\left( \frac{2\pi i}k \right)^nX_n(u) \right\rangle_0}} \nonumber \\
&=& 1+\frac{2\pi i}k\langle W_1(u)\rangle_0+\left( \frac{2\pi i}k \right)^2\Bigl( \langle W_2(u)\rangle_0+\langle X(u)W_1(u)\rangle_0-\langle X(u)\rangle_0\langle W_1(u)\rangle_0 \Bigr)+O(k^{-3}) \nonumber \\ 
   \label{W-CS_2}
\end{eqnarray}
where $\langle{\cal O}(u)\rangle_0$ is the vev in the Gaussian matrix model defined as 
\begin{equation}
\langle {\cal O}(u)\rangle_0\ :=\ \frac1{Z_0}\int d^Nu\,\exp\left[ -\frac12\sum_{i=1}^N(u_i)^2 \right]\prod_{i=1}^N(u_i-u_j)^2\cdot {\cal O}(u), 
\end{equation}
and $W_n(u)$ are defined as 
\begin{eqnarray}
W_m(u) &:=& \frac1N\sum_{i=1}^N\frac{(u_i)^{2m}}{(2m)!}. 
\end{eqnarray}
The vevs in (\ref{W-CS_2}) can be calculated exactly by using the Hermite polynomials. 
The results are as follows: 
\begin{equation}
\begin{array}{rclcrcl}
\langle W_1(u)\rangle_0 &=& \displaystyle{\frac N2}, &\hspace{5mm}& \langle X_1(u)\rangle_0 &=& \displaystyle{\frac{N(N^2-1)}{12}}, \\ [3mm]
\langle W_2(u)\rangle_0 &=& \displaystyle{\frac{2N^2+1}{24}}, &\hspace{5mm}& \langle X_1(u)W_1(u) \rangle_0 &=& \displaystyle{\frac{(N^2-1)(N^2+2)}{24}}. 
\end{array}
\end{equation}
Therefore, the perturbative expansion of the vev $\langle W\rangle$ is given as 
\begin{eqnarray}
\langle W\rangle 
&=& 1+\frac{\pi iN}k+\frac16\left( \frac{2\pi iN}k \right)^2\left( 1-\frac1{4N^2} \right)+O(k^{-3}). 
   \label{expansion-pureCS}
\end{eqnarray}
Note that this is exact in $N$. 
This reproduces the first three terms of the exact result \cite{Kapustin:2009kz}
\begin{equation}
\frac1Ne^{\pi iN/k}\frac{\sin\frac{\pi N}k}{\sin\frac\pi k}\ =\ 1+\frac{\pi iN}k+\frac16\left( \frac{2\pi iN}k \right)^2\left( 1-\frac1{4N^2} \right)+\frac1{24}\left( \frac{2\pi iN}k \right)^3\left( 1-\frac1{2N^2} \right)+O(k^{-4}). 
\end{equation}

\vspace{5mm}

The perturbative calculation described above can be easily extended for the application to the Chern-Simons-matter matrix models with two nodes \cite{Kapustin:2009kz}. 
For these models, we use vevs of the two non-interacting Gaussian matrix models whose partition function is defined as 
\begin{equation}
Z_0\ :=\ \int d^{N_1}ud^{N_2}w\,\exp\left[ -\frac12\sum_{i=1}^{N_1}(u_i)^2-\frac12\sum_{a=1}^{N_2}(w_a)^2 \right]\prod_{i<j}^{N_1}(u_i-u_j)^2\prod_{a<b}^{N_2}(w_a-w_b)^2. 
\end{equation}
The vev $\langle W_1\rangle$ of the Wilson loop for the ${\rm U}(N_1)$ gauge field is given as 
\begin{eqnarray}
\langle W_1\rangle 
&=& 1+\frac{2\pi i}{k_1}\langle W_1(u)\rangle_0+\left( \frac{2\pi i}{k_1} \right)^2\langle W_2(u)\rangle_0+\left( \frac{2\pi i}{k_1} \right)^2\Bigl( \langle Y_1(u)W_1(u)\rangle_0-\langle Y_1(u)\rangle_0\langle W_1(u)\rangle_0 \Bigr) \nonumber \\
& & +\frac{2\pi i}{k_1}\frac{2\pi i}{k_2}\Bigl( \langle Y_2(w)W_1(u)\rangle_0-\langle Y_2(w)\rangle_0\langle W_1(u)\rangle_0 \Bigr)+O(k^{-3}) 
   \label{W-vev}
\end{eqnarray}
where $Y_1(u)$ and $Y_2(w)$, defined as 
\begin{eqnarray}
Y_1(u) &:=&  \frac13\sum_{i<j}^{N_1}\left( \frac{u_i-u_j}2 \right)^2-\frac n2N_2\sum_{i=1}^{N_1}\frac{u_i^2}{4}, \\  
Y_2(w) &:=&  \frac13\sum_{a<b}^{N_2}\left( \frac{w_a-w_b}2 \right)^2-\frac n2N_1\sum_{a=1}^{N_2}\frac{w_a^2}{4},
\end{eqnarray}
come from the one-loop part of the integrand in (\ref{Z-2node}). 
The values of the vevs in (\ref{W-vev}) are 
\begin{equation}
\begin{array}{rclcrcl}
\langle W_1(u)\rangle_0 &=& \displaystyle{\frac{N_1}2}, &\hspace{5mm}& \langle Y_1(u)\rangle_0 &=& \displaystyle{\frac{N_1(N_1^2-1)}{12}-\frac n8N_1^2N_2}, \\ [3mm] 
\langle W_2(u)\rangle_0 &=& \displaystyle{\frac{2N_1^2+1}{24}}, &\hspace{5mm}& \langle Y_1(u)W_1(u)\rangle_0 &=& \displaystyle{\frac{(N_1^2-1)(N_1^2+2)}{24}-\frac n{16}N_1N_2(N_1^2+2)}, \\ [3mm] 
\langle Y_2(w)W_1(u)\rangle_0 &=& \langle Y_2(w)\rangle_0\langle W_1(u)\rangle_0. 
\end{array}
\end{equation}
Therefore, the perturbative expansion of the vev $\langle W_1\rangle$ is given as 
\begin{equation}
\langle W_1\rangle\ =\ 1+\frac{\pi iN_1}{k_1}+\left( \frac{2\pi i}{k_1} \right)^2\left( \frac{4N_1^2-1}{24}-\frac n8N_1N_2 \right)+O(k^{-3}), 
\end{equation}
When $k_1=k, N_1=N_2=N$ and $n=2$, this reproduces the result in \cite{Kapustin:2009kz}. 
Since $(N_1,k_1)$ and $(N_2,k_2)$ appear in the partition function (\ref{Z-2node}) symmetrically, the vev $\langle W_2\rangle$ of the Wilson loop for the ${\rm U}(N_2)$ gauge field must be 
\begin{equation}
\langle W_2\rangle\ =\ 1+\frac{\pi iN_2}{k_2}+\left( \frac{2\pi i}{k_2} \right)^2\left( \frac{4N_2^2-1}{24}-\frac n8N_1N_2 \right)+O(k^{-3}). 
\end{equation}
In terms of the parameters (\ref{parameters-2node}), the vevs can be written as 
\begin{eqnarray}
\langle W_1\rangle 
&=& 1+\frac{t_1}{2\kappa_1}+\frac16\left( \frac{t_1}{\kappa_1} \right)^2\left( 1-\frac1{4N_1^2} \right)-\frac n8\frac{t_1t_2}{\kappa_1^2}+O(t^3), 
   \label{expansion-2node1}\\
\langle W_2\rangle 
&=& 1+\frac{t_2}{2\kappa_2}+\frac16\left( \frac{t_2}{\kappa_2} \right)^2\left( 1-\frac1{4N_2^2} \right)-\frac n8\frac{t_1t_2}{\kappa_2^2}+O(t^3). 
   \label{expansion-2node2}
\end{eqnarray}

In the following, we will show that the planar limit of these expansions can be derived from the planar resolvent obtained in section \ref{2-node}. 

\vspace{5mm}

\subsection{Expansion from the planar resolvent: pure Chern-Simons theory}

\vspace{5mm}

To illustrate how to derive the perturbative expansion from the planar resolvent, let us start with the calculation for pure Chern-Simons theory. 
Recall that the resolvent $v(z)$ satisfies 
\begin{equation}
zv'(z)\ =\ 1-\frac{z-1}{\sqrt{(z-a)(z-b)}}, 
\end{equation}
where $ab=1$ is assumed. 
The 't~Hooft coupling $t$ and the vev $\langle W\rangle$ of the Wilson loop are given as 
\begin{equation}
t\ =\ \frac12\int_C\frac{dz}{2\pi i}\frac{\log z}{z}\frac{z-1}{\sqrt{(z-a)(z-b)}}, \hspace{1cm} t\langle W\rangle\ =\ \frac{a+b-2}4. 
   \label{CS-t and W}
\end{equation}
The vev $\langle W\rangle$ depends on the coupling $t$ through the parameter $a$. 
In order to derive the power series expansion of $\langle W\rangle$ in $t$, it is necessary to know which limit for $a$ corresponds to the weak coupling limit $t\to0$. 

The saddle point equations (\ref{SPE-1}) imply that, for a large $k$ (small $t$), the eigenvalues are expected to be localized around the origin with a narrow width. 
This implies that the limit $t\to0$ corresponds to the limit $a\to1$. 
Introduce a small parameter 
\begin{equation}
\delta\ :=\ -\log a. 
\end{equation}
The expansion in $\delta$ will provide us with the perturbative expansion. 
The integrand in (\ref{CS-t and W}) has the expansion of the following form: 
\begin{equation}
\frac{\log z}{z}\frac{z-1}{\sqrt{(z-a)(z-b)}}\ =\ \sum_{n=0}^\infty f_n(z)\delta^n. 
\end{equation}
Since 
\begin{equation}
\frac{z-1}{\sqrt{(z-a)(z-b)}}\ =\ \left( 1+\frac{1-e^{-\delta}}{z-1} \right)^{-\frac12}\left( 1+\frac{1-e^{\delta}}{z-1} \right)^{-\frac12}, 
\end{equation}
the functions $f_n(z)$ have poles at $z=1$ and are holomorphic elsewhere inside $C$. 
Therefore, the expansion coefficients are given by the residues of $f_n(z)$ at $z=1$. 
Summing up all residues, one obtains 
\begin{equation}
t\ =\ \frac14\delta^2-\frac1{96}\delta^4+\frac1{1440}\delta^6+O(\delta^8). 
\end{equation}
The inverse of this relation is given as 
\begin{equation}
\delta^2\ =\ 4t+\frac23t^2+\frac2{45}t^3+O(t^4). 
\end{equation}
This implies that the perturbative expansion is given as 
\begin{eqnarray}
\langle W\rangle 
&=& \frac1t\left( \frac14\delta^2+\frac1{48}\delta^4+\frac1{1440}\delta^6+O(\delta^8) \right) \nonumber \\
&=& 1+\frac12t+\frac16t^2+O(t^3). 
\end{eqnarray}
This reproduces the planar limit of (\ref{expansion-pureCS}). 

\vspace{5mm}

\subsection{Expansion from the planar resolvent: 2-node theories with $n=2$}

\vspace{5mm}

The perturbative calculation for Chern-Simons-matter theories with 2-node is almost parallel with that for pure Chern-Simons theory shown in the previous subsection, as long as $n=2$. 
Recall that the planar resolvent $\omega(z)$ is 
\begin{equation}
\omega(z)\ =\ -\kappa_2\left[ 1-\frac{z^2-1}{s(z)} \right]-\frac{\kappa_1+\kappa_2}2\frac{z^2-1}{s(z)}\int_{C_1}\frac{d\xi}{2\pi i}\frac{s(\xi)}{\xi(\xi-z)(\xi-z^{-1})}, 
   \label{GT-resolvent}
\end{equation}
where $s(z)$ is defined as (\ref{s(z)}). 
The weak coupling limit $t_1,t_2\to0$ correspond to the limit $\delta_1,\delta_2\to0$ where 
\begin{equation}
\delta_1\ :=\ -\log a_1, \hspace{1cm} \delta_2\ :=\ -\log(-a_2). 
\end{equation}
In this limit, the integral in (\ref{GT-resolvent}) can be evaluated as a power series in $\delta_1$ and $\delta_2$ by evaluating residues at $\xi=1$. 
Then, the 't~Hooft couplings are given in terms of $\delta_1$ and $\delta_2$ as 
\begin{eqnarray}
t_1 
&=& \frac{\kappa_1}4\delta_1^2-\frac{\kappa_1}{96}\delta_1^4+\frac{\kappa_2}{32}\delta_1^2\delta_2^2+\frac{\kappa_1}{1440}\delta_1^6+\frac{6\kappa_1-5\kappa_2}{1536}\delta_1^4\delta_2^2-\frac{5\kappa_2}{1536}\delta_1^2\delta_2^4+O(\delta^8), \\ [2mm]
t_2 
&=& \frac{\kappa_2}4\delta_2^2-\frac{\kappa_2}{96}\delta_2^4+\frac{\kappa_1}{32}\delta_1^2\delta_2^2+\frac{\kappa_2}{1440}\delta_2^6+\frac{6\kappa_2-5\kappa_1}{1536}\delta_1^2\delta_2^4-\frac{5\kappa_1}{1536}\delta_1^4\delta_2^2+O(\delta^8). 
\end{eqnarray}
The inverse of these relations is given as 
\begin{eqnarray}
\delta_1^2 
&=& \frac4{\kappa_1}t_1+\frac{2}{3\kappa_1^2}t_1^2-\frac{2}{\kappa_1^2}t_1t_2+\frac{2}{45\kappa_1^3}t_1^3-\frac{1}{6\kappa_1^3}t_1^2t_2+\frac{\kappa_1+2\kappa_2}{2\kappa_1^3\kappa_2}t_1t_2^2+O(t^4), \\ [2mm]
\delta_2^2 
&=& \frac4{\kappa_2}t_2+\frac{2}{3\kappa_2^2}t_2^2-\frac{2}{\kappa_2^2}t_1t_2+\frac{2}{45\kappa_2^3}t_2^3-\frac{1}{6\kappa_2^3}t_1t_2^2+\frac{2\kappa_1+\kappa_2}{2\kappa_1\kappa_2^3}t_1^2t_2+O(t^4). 
\end{eqnarray}
The linear combination of the vevs of the Wilson loops derived from the expansion of $\omega(z)$ is 
\begin{eqnarray}
t_1\langle W_1\rangle+t_2\langle W_2\rangle 
&=& -\frac{\kappa_2}4\left( e^{-\delta_1}+e^{\delta_1}-e^{-\delta_2}-e^{\delta_2} \right)-\frac{\kappa_1+\kappa_2}4\int_{C_1}\frac{d\xi}{2\pi i}\frac{s(x)}{\xi^2} \nonumber \\
&=& \frac{\kappa_1}{4}{\delta_1}^{2}+\frac{\kappa_2}{4}{\delta_2}^{2}+\frac{\kappa_1}{48}{\delta_1}^{4}+\frac{\kappa_2}{48}{\delta_2}^{4}+\frac{\kappa_1+\kappa_2}{32}{\delta_1}^{2}{\delta_2}^{2}
\nonumber \\
& & +\frac{\kappa_1}{1440}{\delta_1}^{6}+\frac{\kappa_2}{1440}{\delta_2}^{6}+\frac{\kappa_1+\kappa_2}{1536}\delta_1^4\delta_2^2+\frac{\kappa_1+\kappa_2}{1536}\delta_1^2\delta_2^4+O(\delta^8)
\nonumber \\
&=& t_1+t_2+\frac{1}{2\kappa_1}t_1^2+\frac{1}{2\kappa_2}t_2^2+\frac1{6\kappa_1^2}t_1^3-\frac1{4\kappa_1^2}t_1^2t_2-\frac1{4\kappa_2^2}t_1t_2^2
+\frac1{6\kappa_2^2}t_2^3+O(t^4). \nonumber \\
\end{eqnarray}
This reproduces the planar limit of the corresponding linear combination of (\ref{expansion-2node1}) and  (\ref{expansion-2node2}) with $n=2$. 

\vspace{5mm}

\subsection{Expansion from the planar resolvent: 2-node theories with $n\ne2$}

\vspace{5mm}

The planar resolvent for a 2-node theory with $n\ne2$, given in (\ref{resolvent-general}), is quite complicated. 
Indeed, it is given in terms of the theta functions of $u(z)$, and $u(z)$ is given by the inverse of an elliptic function. 
Therefore, the method of calculation used so far in this section does not seem to be appropriate for these general cases. 

A simplification occurs if the range of the parameters is restricted such that $a_1=-a_2=:a$ holds. 
In this case, the quantities $u_0$ and $u_\infty$ can be written simply as 
\begin{equation}
u_0\ =\ \frac14\tau, \hspace{1cm} u_\infty\ =\ \frac14\tau-\frac12. 
\end{equation}
The modulus $\tau$ can be written explicitly as 
\begin{equation}
\tau\ =\ i\frac{2K(a^2)}{K(\sqrt{1-a^4})}. 
\end{equation}
Inverting this relation, one obtains 
\begin{equation}
1-a^4\ =\ 16q^{\frac12}\left( \frac{\sum_{n=1}^\infty q^{\frac12n(n-1)}}{1+2\sum_{n=1}^\infty q^{\frac12n^2}} \right)^4, \hspace{1cm} q\ :=\ e^{\pi i\tau}. 
\end{equation}
As in the previous subsections, we introduce $\delta$ such that $a=\exp(-\delta)$. 
Then, this relation implies that $q^\frac12$ can be given as a power series in $\delta$. 
Explicitly, 
\begin{equation}
q^\frac12\ =\ \frac{1}{4}\delta-\frac{1}{48}\delta^3-\frac{31}{7680}\delta^5+O(\delta^7).  
\end{equation}
This implies that the $q^\frac12$-expansion of the resolvent gives the desired perturbative expansion. 
It turns out that each coefficient of the $q^\frac12$-expansion is a linear combination of exponential functions of $u$. 
Since the 't~Hooft couplings are given as 
\begin{eqnarray}
t_1 
&=& \varphi(b_1)\int_{-\frac12}^{+\frac12}\frac{du}{2\pi i}\frac{\log z(u)}{z(u)}g(u)G(u), 
   \label{t-integral} \\
t_2 
&=& e^{-\pi i\nu}\varphi(b_1)\int_{-\frac12+\frac12\tau}^{+\frac12+\frac12\tau}\frac{du}{2\pi i}\frac{\log z(u)}{z(u)}g(u)G(u),
\end{eqnarray}
the integration of the coefficients can be performed easily. 

To simplify the calculation further, notice that it is enough to perform the perturbative check for $(\kappa_1,\kappa_2)=(1,\pm1)$ since the resolvent for a general $(\kappa_1,\kappa_2)$ is obtained as a linear combination of the resolvents for these two special cases. 

Let us focus on the cases $(\kappa_1,\kappa_2)=(1,\epsilon)$ with $\epsilon=\pm1$. 
The uniqueness of the solution of the saddle point equations (\ref{res-vec1})(\ref{res-vec2}) implies 
\begin{equation}
v_1'(z)\ =\ -\epsilon v_2'(-z). 
\end{equation}
This equality then implies 
\begin{equation}
t_1\ =\ \epsilon t_2\ =:\ t, \hspace{1cm} \langle W_1\rangle\ =\ \langle W_2\rangle\ =:\ \langle W\rangle. 
\end{equation}

The integral formula (\ref{t-integral}) implies 
\begin{eqnarray}
t 
&=& \frac{1}{4}{\delta}^{2}+\epsilon\frac{3{e}^{\pi i\nu}-2\epsilon+3e^{-\pi i\nu}}{192}{\delta}^{4} \nonumber \\
& & +\frac{45{e}^{2\pi i\nu}-150\epsilon{e}^{\pi i\nu}+122-150\epsilon{e}^{-\pi i\nu}+45e^{-2\pi i\nu}}{46080}{\delta}^{6}+O(\delta^8). 
\end{eqnarray}
Inverting this relation, one obtains 
\begin{eqnarray}
\delta^2 
&=& 4t-\epsilon\frac{3{e}^{\pi i\nu}-2\epsilon+3e^{-\pi i\nu}}{3}{t}^{2} \nonumber \\
& & +\frac{45{e}^{2\pi i\nu}+30\epsilon{e}^{\pi i\nu}+98+30\epsilon{e}^{-\pi i\nu}+45e^{-\pi i\nu}}{180}{t}^{3}+O(t^4). 
\end{eqnarray}
The vev $\langle W\rangle$ is given as 
\begin{eqnarray}
(1+\epsilon e^{\pi i\nu})t\langle W\rangle 
&=& -\frac1{4\varphi(b)}\left[ g'(u_0)G(u_0)+g(u_0)G'(u_0) \right] \nonumber \\
&=& \frac{\epsilon{e}^{\pi i\nu}+1}{4}{\delta}^{2}+\frac{3{e}^{2\pi i\nu}+7\epsilon{e}^{\pi i\nu}+7+3\epsilon e^{-\pi i\nu}}{192}{\delta}^{4} \nonumber \\
& & +\epsilon\frac{45{e}^{3\pi i\nu}-15\epsilon{e}^{2\pi i\nu}+62{e}^{\pi i\nu}+62\epsilon-15{e}^{-\pi i\nu}+45\epsilon e^{-2\pi i\nu}}{46080}{\delta}^{6}+O(\delta^8) \nonumber \\
&=& \left( 1+\epsilon {e}^{\pi i\nu}\right) t+\frac{1+\epsilon{e}^{\pi i\nu}}{2}{t}^{2}-\frac{3{e}^{2\pi i\nu}-\epsilon{e}^{\pi i\nu}-1+3\epsilon e^{-\pi i\nu}}{24}{t}^{3}+O(t^4). 
\end{eqnarray}
Therefore, 
\begin{eqnarray}
\langle W\rangle 
&=& 1+\frac12t+\left[ \frac16-\frac n8\epsilon \right]t^2+O(t^3). 
\end{eqnarray}
This reproduces the planar limit of (\ref{expansion-2node1})(\ref{expansion-2node2}). 

\vspace{1cm}

\section{Discussion} \label{discuss}

\vspace{5mm}

We have investigated the planar resolvents of a family of Chern-Simons-matter matrix models which are derived from ${\cal N}\ge3$ Chern-Simons-matter theories with the gauge groups of the form ${\rm U}(N_1)_{k_1}\times{\rm U}(N_2)_{k_2}$ via the supersymmetric localization. 
We found that, although the resolvents themselves are not obtained in general, their derivatives can be determined explicitly. 
From this result, we obtained the explicit formulas for the vevs of the Wilson loops. 
We discussed the possible divergent behaviors of the vevs of the Wilson loops using the explicit formulas. 
As a check of our result, we performed the perturbative calculations of the vevs of Wilson loops. 
The results from the planar resolvents reproduce the results obtained directly from the localization formulas. 

It is interesting to extend the analysis of this paper to a more general family of Chern-Simons-matter matrix models. 
If the gauge group of a given Chern-Simons-matter theory has $g$ factors of ${\rm U}(N)$ type, the resolvent $zv'(z)$ to be determined is valued in a $g$-dimensional vector space with $g$ branch cuts. 
It can be shown that the determination of $zv'(z)$ reduces to a Riemann-Hilbert problem with the monodromy matrices given in terms of the numbers of bi-fundamental hypermultiplets. 
It is interesting to clarify whether some physical quantities like the vevs of the Wilson loops can be obtained in a form explicit enough to investigate their analytic properties. 

We have found for the cases $n>2$ that there exists a strong 't~Hooft coupling limit in which the vevs of the Wilson loops are finite. 
Similar phenomena were also observed in \cite{Suyama:2013fua} for more general theories. 
It would be interesting to analyze the behavior of the physical quantities in the strong coupling limits for the cases $n>2$, and investigate the possibility for the existence of a gravity dual (see e.g. \cite{Billo:2000zr}\cite{Billo:2000zs} for a proposal for the case $n=3$). 

\vspace{2cm}

{\bf \Large Acknowledgements}

\vspace{5mm}

We would like to thank H. Itoyama and T. Oota for valuable discussions. 
This work was supported in part by Fujukai Foundation.

\end{document}